\catcode`\@=11					



\font\fiverm=cmr5				
\font\fivemi=cmmi5				
\font\fivesy=cmsy5				
\font\fivebf=cmbx5				

\skewchar\fivemi='177
\skewchar\fivesy='60


\font\sixrm=cmr6				
\font\sixi=cmmi6				
\font\sixsy=cmsy6				
\font\sixbf=cmbx6				

\skewchar\sixi='177
\skewchar\sixsy='60


\font\sevenrm=cmr7				
\font\seveni=cmmi7				
\font\sevensy=cmsy7				
\font\sevenit=cmti7				
\font\sevenbf=cmbx7				

\skewchar\seveni='177
\skewchar\sevensy='60


\font\eightrm=cmr8				
\font\eighti=cmmi8				
\font\eightsy=cmsy8				
\font\eightit=cmti8				
\font\eightbf=cmbx8				

\skewchar\eighti='177
\skewchar\eightsy='60


\font\ninei=cmmi9
\font\ninesy=cmsy9

\skewchar\ninei='177
\skewchar\ninesy='60


\font\tenrm=cmr10				
\font\teni=cmmi10				
\font\tensy=cmsy10				
\font\tenex=cmex10				
\font\tenit=cmti10				
\font\tensl=cmsl10				
\font\tenbf=cmbx10				
\font\tentt=cmtt10				
\font\tenss=cmss10				
\font\tensc=cmcsc10				
\font\tenbi=cmmib10				

\skewchar\teni='177
\skewchar\tenbi='177
\skewchar\tensy='60

\def\tenpoint{\ifmmode\err@badsizechange\else
	\textfont0=\tenrm \scriptfont0=\sevenrm \scriptscriptfont0=\fiverm
	\textfont1=\teni  \scriptfont1=\seveni  \scriptscriptfont1=\fivemi
	\textfont2=\tensy \scriptfont2=\sevensy \scriptscriptfont2=\fivesy
	\textfont3=\tenex \scriptfont3=\tenex   \scriptscriptfont3=\tenex
	\textfont4=\tenit \scriptfont4=\sevenit \scriptscriptfont4=\sevenit
	\textfont5=\tensl
	\textfont6=\tenbf \scriptfont6=\sevenbf \scriptscriptfont6=\fivebf
	\textfont7=\tentt
	\textfont8=\tenbi \scriptfont8=\seveni  \scriptscriptfont8=\fivemi
	\def\rm{\tenrm\fam=0 }%
	\def\it{\tenit\fam=4 }%
	\def\sl{\tensl\fam=5 }%
	\def\bf{\tenbf\fam=6 }%
	\def\tt{\tentt\fam=7 }%
	\def\ss{\tenss}%
	\def\sc{\tensc}%
	\def\bmit{\fam=8 }%
	\rm\setparameters\setbaselines\fi}


\font\twelverm=cmr12				
\font\twelvei=cmmi12				
\font\twelvesy=cmsy10	scaled\magstep1		
\font\twelveex=cmex10	scaled\magstep1		
\font\twelveit=cmti12				
\font\twelvesl=cmsl12				
\font\twelvebf=cmbx12				
\font\twelvett=cmtt12				
\font\twelvess=cmss12				
\font\twelvesc=cmcsc10	scaled\magstep1		
\font\twelvebi=cmmib10	scaled\magstep1		

\skewchar\twelvei='177
\skewchar\twelvebi='177
\skewchar\twelvesy='60

\def\twelvepoint{\ifmmode\err@badsizechange\else
	\textfont0=\twelverm \scriptfont0=\eightrm \scriptscriptfont0=\sixrm
	\textfont1=\twelvei  \scriptfont1=\eighti  \scriptscriptfont1=\sixi
	\textfont2=\twelvesy \scriptfont2=\eightsy \scriptscriptfont2=\sixsy
	\textfont3=\twelveex \scriptfont3=\tenex   \scriptscriptfont3=\tenex
	\textfont4=\twelveit \scriptfont4=\eightit \scriptscriptfont4=\sevenit
	\textfont5=\twelvesl
	\textfont6=\twelvebf \scriptfont6=\eightbf \scriptscriptfont6=\sixbf
	\textfont7=\twelvett
	\textfont8=\twelvebi \scriptfont8=\eighti  \scriptscriptfont8=\sixi
	\def\rm{\twelverm\fam=0 }%
	\def\it{\twelveit\fam=4 }%
	\def\sl{\twelvesl\fam=5 }%
	\def\bf{\twelvebf\fam=6 }%
	\def\tt{\twelvett\fam=7 }%
	\def\ss{\twelvess}%
	\def\sc{\twelvesc}%
	\def\bmit{\fam=8 }%
	\rm\setparameters\setbaselines\fi}


\font\fourteenrm=cmr12	scaled\magstep1		
\font\fourteeni=cmmi12	scaled\magstep1		
\font\fourteensy=cmsy10	scaled\magstep2		
\font\fourteenex=cmex10	scaled\magstep2		
\font\fourteenit=cmti12	scaled\magstep1		
\font\fourteensl=cmsl12	scaled\magstep1		
\font\fourteenbf=cmbx12	scaled\magstep1		
\font\fourteentt=cmtt12	scaled\magstep1		
\font\fourteenss=cmss12	scaled\magstep1		
\font\fourteensc=cmcsc10 scaled\magstep2	
\font\fourteenbi=cmmib10 scaled\magstep2	

\skewchar\fourteeni='177
\skewchar\fourteenbi='177
\skewchar\fourteensy='60

\def\fourteenpoint{\ifmmode\err@badsizechange\else
	\textfont0=\fourteenrm \scriptfont0=\tenrm \scriptscriptfont0=\sevenrm
	\textfont1=\fourteeni  \scriptfont1=\teni  \scriptscriptfont1=\seveni
	\textfont2=\fourteensy \scriptfont2=\tensy \scriptscriptfont2=\sevensy
	\textfont3=\fourteenex \scriptfont3=\tenex \scriptscriptfont3=\tenex
	\textfont4=\fourteenit \scriptfont4=\tenit \scriptscriptfont4=\sevenit
	\textfont5=\fourteensl
	\textfont6=\fourteenbf \scriptfont6=\tenbf \scriptscriptfont6=\sevenbf
	\textfont7=\fourteentt
	\textfont8=\fourteenbi \scriptfont8=\tenbi \scriptscriptfont8=\seveni
	\def\rm{\fourteenrm\fam=0 }%
	\def\it{\fourteenit\fam=4 }%
	\def\sl{\fourteensl\fam=5 }%
	\def\bf{\fourteenbf\fam=6 }%
	\def\tt{\fourteentt\fam=7}%
	\def\ss{\fourteenss}%
	\def\sc{\fourteensc}%
	\def\bmit{\fam=8 }%
	\rm\setparameters\setbaselines\fi}


\font\seventeenrm=cmr10 scaled\magstep3		


\newdimen\rp@
\newcount\@basestretchnum
\newskip\@baseskip
\newskip\headskip
\newskip\footskip


\def\setparameters{\rp@=.1em
	\headskip=24\rp@
	\footskip=\headskip
	\delimitershortfall=5\rp@
	\nulldelimiterspace=1.2\rp@
	\scriptspace=0.5\rp@
	\abovedisplayskip=10\rp@ plus3\rp@ minus5\rp@
	\belowdisplayskip=10\rp@ plus3\rp@ minus5\rp@
	\abovedisplayshortskip=5\rp@ plus2\rp@ minus4\rp@
	\belowdisplayshortskip=10\rp@ plus3\rp@ minus5\rp@
	\normallineskip=\rp@
	\lineskip=\normallineskip
	\normallineskiplimit=0pt
	\lineskiplimit=\normallineskiplimit
	\jot=3\rp@
	\setbox0=\hbox{\the\textfont3 B}\p@renwd=\wd0
	\skip\footins=12\rp@ plus3\rp@ minus3\rp@
	\skip\topins=0pt plus0pt minus0pt}


\def\setbaselines{\maxdepth=4\rp@\baselinestretch=\@basestretchnum}


\def\baselinestretch{\afterassignment\@basestretch\@basestretchnum}
\def\@basestretch{%
	\@baseskip=12\rp@ \divide\@baseskip by1000
	\normalbaselineskip=\@basestretchnum\@baseskip
	\baselineskip=\normalbaselineskip
	\bigskipamount=\the\baselineskip
		plus.25\baselineskip minus.25\baselineskip
	\medskipamount=.5\baselineskip
		plus.125\baselineskip minus.125\baselineskip
	\smallskipamount=.25\baselineskip
		plus.0625\baselineskip minus.0625\baselineskip
	\setbox\strutbox=\hbox{\vrule height.708\baselineskip
		depth.292\baselineskip width0pt }}



\def\makeheadline{\vbox to0pt{\baselinestretch=1000
	\vskip-\headskip \vskip1.5pt
	\line{\vbox to\ht\strutbox{}\the\headline}\vss}\nointerlineskip}

\def\makefootline{\baselineskip=\footskip\line{\the\footline}}

\def\big#1{{\hbox{$\left#1\vbox to8.5\rp@ {}\right.\n@space$}}}
\def\Big#1{{\hbox{$\left#1\vbox to11.5\rp@ {}\right.\n@space$}}}
\def\bigg#1{{\hbox{$\left#1\vbox to14.5\rp@ {}\right.\n@space$}}}
\def\Bigg#1{{\hbox{$\left#1\vbox to17.5\rp@ {}\right.\n@space$}}}


\mathchardef\alpha="710B
\mathchardef\beta="710C
\mathchardef\gamma="710D
\mathchardef\delta="710E
\mathchardef\epsilon="710F
\mathchardef\zeta="7110
\mathchardef\eta="7111
\mathchardef\theta="7112
\mathchardef\iota="7113
\mathchardef\kappa="7114
\mathchardef\lambda="7115
\mathchardef\mu="7116
\mathchardef\nu="7117
\mathchardef\xi="7118
\mathchardef\pi="7119
\mathchardef\rho="711A
\mathchardef\sigma="711B
\mathchardef\tau="711C
\mathchardef\upsilon="711D
\mathchardef\phi="711E
\mathchardef\chi="711F
\mathchardef\psi="7120
\mathchardef\omega="7121
\mathchardef\varepsilon="7122
\mathchardef\vartheta="7123
\mathchardef\varpi="7124
\mathchardef\varrho="7125
\mathchardef\varsigma="7126
\mathchardef\varphi="7127
\mathchardef\imath="717B
\mathchardef\jmath="717C
\mathchardef\ell="7160
\mathchardef\wp="717D
\mathchardef\partial="7140
\mathchardef\flat="715B
\mathchardef\natural="715C
\mathchardef\sharp="715D


\def\err@badsizechange{%
	\immediate\write16{--> Size change not allowed in math mode, ignored}}

\baselinestretch=1000
\tenpoint

\catcode`\@=12					
\catcode`\@=11
\expandafter\ifx\csname @iasmacros\endcsname\relax
	\global\let\@iasmacros=\par
\else	\immediate\write16{}
	\immediate\write16{Warning:}
	\immediate\write16{You have tried to input iasmacros more than once.}
	\immediate\write16{}
	\endinput
\fi
\catcode`\@=12


\def\rmb{\seventeenrm}

\def\singlespace{\baselineskip=\normalbaselineskip}

\def\doublespace{\baselineskip=2\normalbaselineskip}


\def\nonarrower{\advance\leftskip by-\parindent
	\advance\rightskip by-\parindent}


\def\boxit#1{\vbox{\hrule\hbox{\vrule\kern3pt
	\vbox{\kern3pt#1\kern3pt}\kern3pt\vrule}\hrule}}

\def\hence{\leavevmode\hbox{\bf .\raise5.5pt\hbox{.}.} }

\def\dalemb#1#2{{\vbox{\hrule height.#2pt
	\hbox{\vrule width.#2pt height#1pt \kern#1pt \vrule width.#2pt}
	\hrule height.#2pt}}}
\def\gtorder{\mathrel{\raise.3ex\hbox{$>$}\mkern-14mu
             \lower0.6ex\hbox{$\sim$}}}
\def\ltorder{\mathrel{\raise.3ex\hbox{$<$}\mkern-14mu
             \lower0.6ex\hbox{$\sim$}}}

\newdimen\fullhsize
\newbox\leftcolumn
\def\twoup{\hoffset=-.5in \voffset=-.25in
  \hsize=4.75in \fullhsize=10in \vsize=6.9in
  \def\fullline{\hbox to\fullhsize}
  \let\lr=L
  \output={\if L\lr
        \global\setbox\leftcolumn=\columnbox\global\let\lr=R \advancepageno
      \else \doubleformat \global\let\lr=L\fi
    \ifnum\outputpenalty>-20000 \else\dosupereject\fi}
  \def\doubleformat{\shipout\vbox{
    \fullline{\box\leftcolumn\hfil\columnbox}\advancepageno}}
  \def\columnbox{\leftline{\vbox{\makeheadline\pagebody\makefootline}}}
  \tolerance=1000 }
\twelvepoint
\doublespace
{\nopagenumbers{
\rightline{~~~November, 2004}
\bigskip\bigskip
\centerline{\rmb \fourteenpoint ANOMALIES}
\medskip
\centerline{\bf Stephen L. Adler\footnote{$^*$}{\rm \tenpoint
Work supported in part by the Department of Energy under
Grant \#DE--FG02--90ER40542.} }

\centerline{\bf Institute for Advanced Study}
\centerline{\bf Princeton, NJ 08540}
\medskip
\bigskip\bigskip
\leftline{\it Send correspondence to:}
\medskip
{\singlespace\leftline{Stephen L. Adler}
\leftline{Institute for Advanced Study}
\leftline{Einstein Drive, Princeton, NJ 08540}
\leftline{Phone 609-734-8051; FAX 609-924-8399; email adler@ias.
edu}}
\bigskip\bigskip
}}
\vfill\eject
\pageno=2
\bigskip
\leftline{\bf Synopsis} 
\bigskip

{\bf Anomalies} are the breaking of classical symmetries by quantum 
mechanical radiative corrections, which arise when the  
regularizations needed to evaluate small fermion loop 
{}Feynman diagrams conflict with a classical symmetry of the theory.  
They have important implications for a wide range of issues in quantum 
field theory, mathematical physics, and string theory. 

\bigskip
\leftline{\bf Chiral Anomalies, Abelian and Non-Abelian}
\bigskip
Consider quantum electrodynamics, with the Lagrangian density 
$${\cal L}=\overline \psi( i \gamma^{\mu}\partial_{\mu}
-e_0  \gamma^{\mu}B_{\mu}-m_0) \psi     ~~~,\eqno[1a]$$
where $\overline \psi= \psi^{\dagger} \gamma^0$, where $e_0$ and $m_0$   
are the bare charge and mass, and where $B_{\mu}$ is the electromagnetic  
gauge potential.  (We reserve the notation $A$ for axial-vector quantities.)  
Under a chiral transformation 
$$\psi \to e^{i \lambda \gamma_5} \psi ~~~,\eqno[1b] $$ 
with constant $\lambda$, the kinetic term in eqn [1a] is invariant (because 
$\gamma_5$ commutes with $\gamma^0 \gamma^{\mu}$), whereas the mass term is 
not invariant.  Therefore, naive application of Noether's theorem would lead 
one to expect that the axial-vector current 
$$j_{\mu}^5=\overline \psi \gamma_{\mu} \gamma_5 \psi~~~, \eqno[1c]$$ 
obtained from the Lagrangian density by applying a chiral transformation 
with spatially varying $\lambda$, should have a divergence given by the 
change under chiral transformation of the 
mass term in eqn [1a].  To tree approximation, this is indeed true, but 
when one computes the $AVV$ Feynman diagram with one axial-vector and two 
vector vertices (see {\bf Figure 1}), and insists on 
conservation of the vector current 
$j_{\mu}=\overline\psi \gamma_{\mu} \psi$, one finds 
that to order $e_0^2$ the classical Noether 
theorem is modified to read 
$$\partial^{\mu}j_{\mu}^5(x)=2 i m_0 j^5(x) + {e_0^2\over 16 \pi^2} 
{}F^{\xi\sigma}(x) F^{\tau\rho}(x) \epsilon_{\xi\sigma\tau\rho}~~~,\eqno[2]$$
with $F^{\xi\sigma}(x)=\partial^{\sigma}B^{\xi}(x)
-\partial^{\xi}B^{\sigma}(x)  $ the electromagnetic field-strength tensor.  
The second term 
in eqn [2], which would be unexpected from application of the 
classical Noether theorem, is the Abelian axial-vector anomaly, often called 
the Adler-Bell-Jackiw (or ABJ) anomaly after the seminal papers on 
the subject. Since vector current conservation together with the axial-vector 
current anomaly implies that the left and right handed 
chiral currents $j_{\mu}\pm j_{\mu}^5$ are also anomalous, the axial-vector  
anomaly is frequently called the {\it chiral anomaly}, and we shall use the 
terms interchangeably in this article.  

There are a number of different ways to understand why the extra term 
in eqn [2] appears.  
(1)  If one works through the formal Feynman diagrammatic 
Ward identity proof of the Noether theorem, one finds that there is a step 
where the closed fermion loop contributions are eliminated by 
a shift of the loop integration variable.  For Feynman diagrams that are  
convergent this is no problem, but the $AVV$ diagram is linearly divergent.  
The linear divergence vanishes under symmetric integration, but the shift 
then produces a finite residue, which gives the anomaly. (2)  If one defines 
the $AVV$ diagram by Pauli-Villars regularization with regulator mass 
$M_0$ that is allowed to approach infinity at the end of the calculation, 
one finds a classical Noether theorem in the regulated theory, 
$$\partial^{\mu} j_{\mu}^5|_{m_0} -\partial^{\mu} j_{\mu}^5|_{M_0}
=2im_0 j^5|_{m_0}-2iM_0 j^5|_{M_0}~~~,\eqno[3a]$$
with the subscripts $m_0$ and $M_0$ indicating that fermion loops are to 
be calculated with fermion mass $m_0$ and $M_0$ respectively.  Taking the 
vacuum to two photon matrix element of eqn [3a], one finds that the 
the matrix element $\langle 0|j^5|_{M_0}|\gamma \gamma \rangle$, which is 
unambiguously computable after imposing vector current conservation, falls 
off only as $M_0^{-1}$ as the regulator mass approaches infinity. Thus,
the product of $2iM_0$ with this matrix element has a finite limit, which 
gives the anomaly. (3)  If one defines the gauge invariant 
axial-vector current by  
point-splitting, 
$$j_{\mu}^5(x)=\overline \psi(x+\epsilon/2) \gamma_{\mu} \gamma_5 
\psi(x-\epsilon/2) e^{-ie_0\epsilon^{\sigma}B_{\sigma}(x)}~~~,
\eqno[3b]$$
with $\epsilon \to 0$ at the end of the calculation, one finds that 
the divergence of eqn [3b] contains an extra term with  
a factor of $\epsilon$. On careful evaluation one finds that the coefficient 
of this factor is an expression that behaves as $\epsilon^{-1}$,  
which gives the anomaly in the limit of vanishing $\epsilon$.  (4)  Finally,   
if one defines the field theory by a functional integral over the classical 
action, the standard Noether analysis shows that the classical action is 
invariant under the chiral transformation of eqn [1b], apart from the 
contribution of the mass term, which gives the naive axial-vector divergence.  
However, as pointed out by Fujikawa, 
one must also apply the chiral transformation to the functional 
integration measure, and since the measure is an infinite product it must 
be regularized to be well defined.  Careful calculation shows that the 
regularized measure is not chiral invariant, but contributes an extra term 
to the axial-vector Ward identity that is precisely the chiral anomaly.  

A key feature of the anomaly is that it is {\it irreducible}: one cannot add 
any local polynomial counter term to the $AVV$ diagram that preserves vector 
current conservation, and which eliminates the anomaly.  More generally, 
one can show that there is no way of modifying quantum electrodynamics so 
as to eliminate the chiral anomaly, without spoiling either vector current 
conservation (that is, electromagnetic gauge invariance), renormalizability, 
or unitarity.  Thus the chiral anomaly is a new physical effect in 
renormalizable quantum field theory, that is not present in  
the pre-quantization classical theory.  

The Abelian chiral anomaly is the simplest case of the anomaly phenomenon. 
It was extended to non-Abelian gauge theories by Bardeen using a 
point-splitting method to compute the divergence, followed by   
adding polynomial counter terms to remove as many of the residual terms as 
possible.  The resulting irreducible divergence is the non-Abelian chiral 
anomaly, which in terms of Yang-Mills field strengths for vector and 
axial-vector gauge potentials $V^{\mu}$ and $A^{\mu}$, 
$$\eqalign{
F_V^{\mu\nu}(x)=&\partial^{\mu}V^{\nu}(x)-\partial^{\nu}V^{\mu}(x) 
-i[V^{\mu}(x),V^{\nu}(x)]-i[A^{\mu}(x),A^{\nu}(x)]~~~,\cr
F_A^{\mu\nu}(x)=&\partial^{\mu}A^{\nu}(x)-\partial^{\nu}A^{\mu}(x) 
-i[V^{\mu}(x),A^{\nu}(x)]-i[A^{\mu}(x),V^{\nu}(x)]~~~,\cr
}\eqno(4a)$$
is given by 
$$\eqalign{
\partial^{\mu}j_{5\mu}^{a}(x)=&{\rm normal~divergence~term} \cr
+&(1/4\pi^2)\epsilon_{\mu\nu\sigma\tau}{\rm tr} \lambda_A^{a}
[(1/4)F_V^{\mu\nu}(x)F_V^{\sigma \tau}(x)
+(1/12)F_A^{\mu\nu}(x)F_A^{\sigma \tau}(x) ~~~\cr
+&(2/3)iA^{\mu}(x)A^{\nu}(x)F_V^{\sigma \tau}(x) 
+(2/3)i F_V^{\mu \nu}(x) A^{\sigma}(x)A^{\tau}(x) 
+(2/3)i  A^{\mu}(x)F_V^{\nu \sigma}(x) A^{\tau}(x) \cr
-&(8/3) A^{\mu}(x)A^{\nu}(x)A^{\sigma}(x)A^{\tau}(x) ]~~~.\cr
}\eqno[4b]$$ 
In eqn [4b],  ${\rm tr}$ denotes a trace over 
internal degrees of freedom, and   
$\lambda_A^{a}$ is the internal symmetry matrix associated with 
the axial-vector external field.  In the Abelian case, where there is no   
internal symmetry structure, 
the terms involving two or four factors of $A^{\mu},\,A^{\nu},...$ vanish by 
antisymmetry of $\epsilon_{\mu\nu\sigma\tau}$, and one recovers the $AVV$ 
triangle anomaly, as well as a kinematically related anomaly in the 
$AAA$ triangle diagram. In the non-Abelian case, with non-trivial internal  
symmetry structure,  there are also box and pentagon diagram anomalies. 

In addition to coupling to spin 1 gauge fields, fermions can also couple 
to spin 2 gauge fields, associated with the graviton.    
When the coupling of fermions to gravitation is taken into account,  
the axial-vector current $\overline \psi T\gamma_{\mu} \gamma_5 \psi$, 
with $T$ an internal symmetry matrix, has an additional anomalous 
contribution to its divergence proportional to 
$${\rm tr} T \epsilon_{\xi\sigma\tau\rho} R^{\xi\sigma\alpha\beta}
R^{\tau\rho}_{~~\alpha\beta}~~~, \eqno[4c]$$
where $R_{\xi\sigma\tau\rho}$ is the Riemann curvature tensor of the 
gravitational field.  

\bigskip
\leftline{\bf Chiral Anomaly Nonrenormalization}

A salient feature of the chiral anomaly is the fact that it is not 
renormalized by higher order radiative corrections.  In other words, 
the one loop expressions of eqns [2] and [4b] give the {\it exact} 
anomaly coefficient, without modification in higher orders of perturbation 
theory.  In gauge theories such as quantum electrodynamics and quantum 
chromodynamics, this result (the Adler-Bardeen theorem) can be understood 
heuristically as follows.  Write down a modified Lagrangian in which 
regulators are included for all gauge boson fields.  Since the gauge 
boson regulators 
do not influence the chiral symmetry properties of the theory, the 
divergences of the chiral currents are not affected by their inclusion,  
and so the only sources of anomalies in the regularized theory are small 
single-fermion loops, giving the anomaly expressions of eqns [2] and [4b].  
Since the renormalized theory is obtained as the limit of the regularized 
theory as the regulator masses approach infinity, this result applies 
to the renormalized theory as well.  

The above argument can be made precise, and extends to non-gauge theories 
such as the $\sigma$-model as well.  For both gauge theories and the 
$\sigma$-model, cancellation of radiative corrections 
to the anomaly coefficient has been explicitly demonstrated in 
fourth order calculations.  Nonperturbative demonstrations of anomaly 
renormalization have also been given using the Callan-Symanzik equations. 
{}For example, in quantum electrodynamics, Zee,  and Lowenstein and Schroer,   
showed that a factor $f$ that 
gives the {\it ratio} of the true anomaly to its one-loop value obeys 
the differential equation 
$$(m{\partial \over \partial m}+ 
\alpha \beta(\alpha) {\partial \over \partial \alpha})
 f=0~~~.\eqno[5]$$
Since $f$ is dimensionless it can have no dependence on the mass $m$, and 
since $\beta(\alpha)$ is nonzero this implies 
$\partial f/ \partial \alpha=0$. Thus $f$ has no dependence on 
$\alpha$, and so $f=1$.  

\bigskip
\leftline{\bf Applications of Chiral Anomalies}

Chiral anomalies have numerous applications in the standard model of 
particle physics and its extensions, and we describe here a few of the 
most important ones. 
\medskip
\noindent
(1)  {\it Neutral pion decay $\pi^0 \to \gamma \gamma$}.     
As a result of the Abelian chiral anomaly, the partially conserved 
axial-vector current (PCAC) equation relevant to neutral pion decay is 
modified to read 
$$\partial^{\mu} {\cal F}_{3\mu}^5(x)= (f_{\pi}\mu_{\pi}^2  
/\sqrt{2}) \phi_{\pi}(x)
+S {\alpha_0 \over 4 \pi} 
F^{\xi \sigma}(x) F^{\tau \rho}(x) \epsilon_{\xi\sigma \tau \rho}
~~~,\eqno[6a]$$   
with $\mu_{\pi}$  the pion mass, $f_{\pi}\simeq 131\,$MeV 
the charged pion decay constant,
and with $S$ a constant determined by the constituent fermion charges and 
axial-vector couplings.  Taking the matrix element of eqn [6a] between the 
vacuum state and a two photon state, and using the fact that the left hand 
side has a kinematic zero (the Sutherland-Veltman theorem), one sees that 
the  $\pi^0 \to \gamma \gamma$ amplitude $F$ is completely determined by 
the anomaly term, giving the formula 
$$F=-(\alpha/\pi) 2S \sqrt{2} /f_{\pi}~~~.\eqno[6b]$$
{}For a single set of fractionally charged quarks,  the amplitude 
$F$ is a factor of three too small to agree with experiment; for three 
fractionally charged quarks (or an equivalent Han-Nambu triplet), eqn [6b]   
gives the correct neutral pion decay rate.  This calculation was one of 
the first pieces of evidence for the color degree of freedom of quarks.  
\medskip
\noindent
(2)  {\it Anomaly cancellation in gauge theories}.  In quantum 
electrodynamics 
the gauge particle (the photon)  couples to the vector current, and so 
the anomalous conservation properties of the axial-vector current have no 
effect.  The same statement holds for the gauge gluons in quantum 
chromodynamics, when treated in isolation from the other interactions.    
However, in the electroweak theory that embeds quantum electrodynamics in 
a theory of the weak force, the gauge particles (the $W^{\pm}$ and $Z$ 
intermediate bosons) couple to chiral currents, which are left or right 
handed linear combinations of the vector and axial-vector currents. In 
this case, the chiral anomaly leads to problems with the renormalizability 
of the theory, unless the anomalies cancel between different fermion 
species.  Writing all fermions as left handed, the condition for anomaly 
cancellation is 
$$
{\rm tr} \{T_{\alpha},T_{\beta}\}T_{\gamma}= 
{\rm tr} (T_{\alpha}T_{\beta}+T_{\beta}T_{\alpha})T_{\gamma}=0~,    
~~~{\rm all~~} \alpha,\,\beta,\,\gamma~~~,
\eqno[7]$$
with $T_{\alpha}$ the coupling matrices of gauge bosons to left handed 
fermions.  These conditions are obeyed in the standard model, by virtue 
of three non-trivial sum rules on the fermion gauge couplings being 
satisfied (four sum rules if one includes the 
gravitational contribution to the 
chiral anomaly, given in eqn [4c], which also cancels in the standard model.)  
Note that anomaly cancellation in the {\it locally} gauged currents of the 
standard model does not 
imply anomaly cancellation in {\it global} flavor currents.  Thus the 
flavor axial-vector current anomaly that gives the $\pi^0 \to \gamma \gamma$ 
matrix element remains anomalous in the full electroweak theory.    
Anomaly cancellation imposes 
important constraints on the construction of grand unified models that 
combine the electroweak theory with quantum chromodynamics.  For instance, 
in $SU(5)$ the fermions are put into a $\overline 5$ and $10$ representation, 
which together, but not individually, are anomaly free. The larger 
unification groups $SO(10)$ and $E_6$ satisfy eqn [7] for 
all representations, 
and so are automatically anomaly free.  
\medskip
\noindent
(3) {\it Instanton physics and the theta vacuum}. The theory of anomalies  
is intimately tied to the physics associated with instanton classical
Yang-Mills theory solutions.  Since the instanton field strength is 
self-dual, the nonvanishing instanton Euclidean action 
$$S_E=\int d^4 x {1\over 4} F_{\mu\nu}F^{\mu\nu}=8\pi^2
~~~\eqno[8a]$$ 
implies that the integral 
of the pseudoscalar density 
$F_{\mu\nu}F_{\lambda \sigma}\epsilon^{\mu\nu\lambda\sigma}$ over the 
instanton is also nonzero, 
$$\int d^4 x F_{\mu\nu}F_{\lambda \sigma}
\epsilon^{\mu\nu\lambda\sigma}   =64\pi^2  ~~~.\eqno[8b]$$
Referring back to eqn [4b], we see that this means that the integral of 
the non-Abelian chiral anomaly for fermions in the background field of  
an instanton is an integer, which in the Minkowski space continuation  
has the interpretation of a topological winding  number change produced 
by the instanton tunneling solution.  This fact has a number of profound 
consequences.  Since a vacuum with definite winding number $|\nu\rangle$ 
is unstable 
under instanton tunneling, careful analysis shows that the 
non-Abelian vacuum that has correct clustering properties is a Fourier 
superposition  
$$|\theta\rangle =\sum_{\nu} e^{i\theta \nu} |\nu\rangle~~~,\eqno[8c]$$
giving rise to the $\theta$-vacuum of quantum chromodynamics, and a 
host of issues associated with (the lack of) strong $CP$-violation, 
the Peccei-Quinn mechanism, and axion 
physics.  Also, the fact that the integral of eqn [8b] is  
nonzero means that the $U(1)$ chiral symmetry of quantum chromodynamics is 
broken by instantons, which as shown by 't Hooft resolves the longstanding 
``$U(1)$ problem'' of the strong interactions, that of explaining why  the 
flavor singlet pseudoscalar meson $\eta^{\prime}$ is not light, as are 
its flavor octet partners. 
\medskip
\noindent
(4)  {\it Anomaly matching conditions}.  The anomaly structure of a theory, 
as shown by 't Hooft, leads to important constraints on the formation of 
massless composite bound states.  Consider a theory with a set of left handed 
fermions $\psi^{if}$, with $i$ a ``color'' index acted on by a 
non-Abelian gauge force, and $f$ an ungauged family or ``flavor'' index.  
Suppose that the family multiplet structure is such that the global chiral 
symmetries associated with the flavor index have nonvanishing anomalies 
${\rm tr} \{T_{\alpha},T_{\beta}\}T_{\gamma}$.  Then the 't Hooft condition     
asserts that if the color forces result in the formation of composite 
massless bound states of the original completely confined fermions,  
and if there is no spontaneous breaking of the original 
global flavor symmetries, then these bound states 
must contain left handed spin-1/2 composites with 
a representation structure $S$ that has the same anomaly coefficient as 
that in the underlying theory. In other words, we must have 
$${\rm tr} \{S_{\alpha},S_{\beta}\}S_{\gamma} =
{\rm tr} \{T_{\alpha},T_{\beta}\}T_{\gamma}  ~~~.\eqno[9]$$
To prove this, one adjoins to the theory a set of right-handed 
spectator fermions 
$\psi^{f}$  with the same flavor structure as the original set, but which 
are not acted on by the color force.  These right handed fermions 
cancel the original anomaly, making the underlying theory at zero 
color coupling anomaly-free; 
since dynamics cannot spontaneously generate anomalies, the 
theory when 
the color dynamics is turned on must also have 
no global chiral anomalies. This  
implies that the bound state spectrum must conspire to cancel the anomalies 
associated with the right-handed spectators, in other words, the bound state  
anomaly structure must match that of the original fermions.  This anomaly 
matching condition has found applications in the study of the 
possible compositeness of quarks and leptons.  It has also been 
applied to the 
derivation of non-perturbative dynamical results in whole classes of  
supersymmetric theories,  where the combined tools  of holomorphicity, 
instanton physics, and anomaly matching have given incisive 
results.  

\bigskip
\leftline{\bf Global Structure of Anomalies}
\bigskip
We noted above that the chiral anomalies are irreducible, in that they
cannot be eliminated by adding a local polynomial counter-term to the 
action.  However, anomalies can be described by a non-local effective 
action, obtained by integrating out the fermion field dynamics, 
and this point of view proves very useful in the non-Abelian case. 
Starting with the Abelian case for orientation, we note that if $A^{\mu}$ is 
an external axial-vector field, and we write an effective action 
$\Gamma[A]$, then the axial-vector current $j_{\mu}^5$ associated with 
$A^{\mu}$ is given (up to an overall constant) by the variational 
derivative expression 
$$j_{\mu}^5(x)= {\delta \Gamma[A] \over \delta A^{\mu}(x) }  ~~~,\eqno[10a]$$
and the Abelian anomaly appears as the fact that the expression  
$$\eqalign{
\partial^{\mu}j_{\mu}^5=&X\Gamma[A] =G \not= 0~~~, \cr
X=&\partial^{\mu} {\delta \over \delta  A^{\mu}(x) } ~~~,\cr
}\eqno[10b]~~~$$
is non-vanishing even when the theory is classically  
chiral invariant.  Turning now to the non-Abelian 
case, the variational derivative appearing in eqns [10a,b] must be replaced 
by an appropriate covariant derivative.   In terms of the internal 
symmetry component fields $A_{\mu}^a$ and $V_{\mu}^a$ of the Yang-Mills 
potentials of eqn [4a], one introduces operators 
$$\eqalign{
-X^a(x)=&\partial^{\mu}{\delta \over \delta A_{\mu}^a(x)} 
+f_{abc}V_{\mu}^b{\delta \over \delta A_{\mu}^c(x)} 
+f_{abc}A_{\mu}^b{\delta \over \delta V_{\mu}^c(x)}~~~,\cr
-Y^a(x)=&\partial^{\mu}{\delta \over \delta V_{\mu}^a(x)} 
+f_{abc}V_{\mu}^b{\delta \over \delta V_{\mu}^c(x)} 
+f_{abc}A_{\mu}^b{\delta \over \delta A_{\mu}^c(x)}~~~,\cr
}\eqno[11a]$$
with $f_{abc}$ the antisymmetric non-Abelian group structure constants.   
The operators $X^a$ and $Y^a$ are easily seen to obey the 
commutation relations
$$\eqalign{
[X^a(x),X^b(y)]=&  f_{abc}\delta(x-y) Y_c(x)~~~,\cr  
[X^a(x),Y^b(y)]=&  f_{abc}\delta(x-y) X_c(x)~~~,\cr  
[Y^a(x),Y^b(y)]=&  f_{abc}\delta(x-y) Y_c(x)~~~.\cr  
}\eqno[11b]$$
Let $\Gamma[V,A]$ be the effective action as a functional of the 
fields $V^{\mu},\,A^{\mu}$, constructed so that the vector 
currents are covariantly 
conserved, as expressed formally by 
$$Y^a \Gamma[V,A]=0~~~.\eqno[12a]$$
Then the non-Abelian axial-vector current anomaly is given by 
$$X^a \Gamma[V,A]=G^a~~~.\eqno[12b]$$
{}From eqns [12a,b] and the first line of eqn [11b], we have 
$$X^b G^a-X^a G^b= (X^bX^a-X^aX^b)\Gamma[V,A]
\propto f_{abc} Y^c \Gamma[V,A] =0~~~,\eqno[12c]$$
which is the Wess-Zumino consistency condition on the 
structure of the anomaly $G^a$.  It can be shown that this condition uniquely 
fixes the  form of the non-Abelian anomaly to be that of eqn [4b], up to an 
overall constant, which can be determined by comparison with the simplest 
anomalous $AVV$ triangle graph.   A physical consequence of the consistency 
condition is that the $\pi^0 \to \gamma \gamma$ decay amplitude determines 
uniquely certain other anomalous amplitudes, such as $2\gamma \to 3 \pi$, 
$\gamma \to 3 \pi$, and a five pseudoscalar vertex.  

Although the action $\Gamma[V,A]$ is 
necessarily nonlocal, Wess and Zumino were able to write down a local 
action, involving an auxiliary pseudoscalar field, that obeys the anomalous 
Ward identities and the consistency conditions.   Subsequently, Witten 
gave a new construction of this local action, in terms of the integral of a
fifth rank antisymmetric tensor over a five dimensional disk which has four
dimensional space as its boundary.  He also showed that requiring 
$e^{i\Gamma}$ to be independent of the choice of the spanning disk requires, 
in analogy with Dirac's quantization condition for monopole charge, the 
condition that the overall coefficient in the non-Abelian anomaly  be 
quantized in integer multiples. Comparison with the lowest order triangle 
diagram shows that in the case of $SU(N_c)$ gauge theory, this integer is 
just the number of colors $N_c$.  Thus, global considerations tightly 
constrain the non-Abelian chiral anomaly structure, and dictate that up 
to an integer proportionality constant, it must have the 
form given in eqns [4a,b].
\bigskip
\leftline{\bf Trace Anomalies}

The discovery of chiral anomalies inspired the search for other examples 
of anomalous behavior.  First indications of a  perturbative trace anomaly 
obtained in a study of broken scale invariance by Coleman and Jackiw were 
shown by Crewther, and by Chanowitz and Ellis, to correspond to an anomaly 
in the three point function $\theta_{\sigma}^{\sigma} V_{\mu}V_{\nu}$, where  
$\theta_{\mu}^{\mu}$ is the energy-momentum tensor.   
Letting $\Delta_{\mu\nu}
(p)$ be  the momentum space expression for this three-point function, and 
$\Pi_{\mu\nu}$  the corresponding $V_{\mu}V_{\nu}$ two-point function, the 
trace anomaly equation in quantum electrodynamics reads 
$$ \Delta_{\mu\nu}(p)=\left(2-p_{\sigma}{\partial\over \partial p_{\sigma}}
\right) \Pi_{\mu\nu}(p)-{R\over 6 \pi^2}(p_{\mu}p_{\nu}-\eta_{\mu\nu}p^2) 
~~~~,\eqno[13a]$$
with the first term on the right hand side the naive divergence, and the 
second term the trace anomaly, with anomaly coefficient $R$ given by 
$$R=\sum_{i,{\rm spin} {1\over 2}}Q_i^2 
+{1\over 4} \sum_{i,{\rm spin} 0}Q_i^2~~~.\eqno[13b]$$ 
The fact that there should be a trace anomaly can readily be inferred from 
a trace analog of the Pauli-Villars regulator argument for the chiral anomaly 
given in eqn [3a]. Letting $j=\overline \psi \psi$ be the scalar 
current in Abelian electrodynamics, one has 
$$\theta_{\mu}^{\mu}|_{m_0} - \theta_{\mu}^{\mu}|_{M_0} = 
m_0 j|_{m_0} - M_0 j|_{M_0}~~~.\eqno[13c]$$
Taking the vacuum to two photon matrix element of this equation, 
and imposing vector current conservation, one 
finds that the matrix element $\langle 0|j|_{M_0}|\gamma \gamma\rangle$ 
is proportional to 
$M_0^{-1}\langle 0| F_{\lambda \sigma} F^{\lambda \sigma} 
|\gamma \gamma \rangle_{M_0}$ for large  
regulator mass, and so makes a non-vanishing contribution to the right hand 
side of eqn [13c], giving the lowest order trace anomaly.  

Unlike the chiral anomaly, the trace anomaly {\it is} renormalized in 
higher orders of perturbation theory; heuristically, the reason is that 
whereas boson field regulators  do not affect the chiral symmetry properties 
of a gauge theory (which are determined just by the fermionic terms in the 
Lagrangian), they do alter the energy-momentum tensor, since gravitation 
couples to all fields, including regulator fields.  An analysis using the 
Callan-Symanzik equations shows, however, that the trace anomaly is 
computable to all orders in terms of various renormalization group functions 
of the coupling. For example, in Abelian electrodynamics, defining 
$\beta(\alpha)$ and $\delta(\alpha)$ by 
$\beta(\alpha)= (m/\alpha) \partial \alpha/
\partial m$ and   
$1+\delta(\alpha)=(m/m_0)\partial m_0/\partial m$, the trace of the 
energy-momentum tensor is given to all orders by 
$$\theta_{\mu}^{\mu} = [1+\delta(\alpha)] m_0 \overline \psi \psi 
+ {1\over 4} \beta(\alpha)  N[F_{\lambda \sigma}F^{\lambda \sigma}]  
+... ~~~,\eqno[14]$$
with $N[...]$ specifying conditions that make the division into 
two terms in eqn [14] unique.  
A similar relation holds in the non-Abelian case, again with the $\beta$ 
function appearing as the coefficient of the anomalous ${\rm tr}   
 N[F_{\lambda \sigma}F^{\lambda \sigma}]$ term.    

Just as in the chiral anomaly case, when spin-0, spin-1/2, or spin-1 
fields propagate on a background spacetime, there are curvature dependent 
contributions to the trace anomaly, in other words, gravitational 
anomalies.   These typically take the form of complicated linear combinations 
of terms of the form $R^2,~ R_{\mu\nu}R^{\mu\nu},~ R_{\mu\nu\lambda\sigma}
R^{\mu\nu\lambda\sigma},~R_{,\mu}^{~~~;\mu}$, with coefficients depending 
on the matter fields involved.  

In supersymmetric theories, the axial-vector current and the energy-momentum 
tensor are both components of the supercurrent, and so their anomalies 
imply the existence of corresponding supercurrent anomalies.  The issue 
of how the non-renormalization of chiral anomalies (which have a supercurrent  
generalization given by the Konishi anomaly), and the renormalization 
of trace anomalies,  can coexist in supersymmetric theories originally 
engendered considerable confusion.  This apparent puzzle 
is now understood in the context of a perturbatively exact expression 
for the $\beta$ function in supersymmetric field theories (the so-called 
NSVZ, for Novikov, Shifman, Vainshtein, and Zakharov,  $\beta$ function).  
Supersymmetry anomalies 
can be used to infer the structure of effective actions in supersymmetric 
theories, and these in turn have important implications for possibilities 
for dynamical supersymmetry breaking.  Anomalies may also play a role, 
through anomaly mediation, in communicating supersymmetry breaking in 
``hidden sectors'' of a theory, that do not contain the physical fields 
that we directly observe, to the ``physical sector'' containing the 
observed fields.
\bigskip
\leftline{\bf Further Anomaly Topics}  
\bigskip
The above discussion has focused on some of the principal features and 
applications of anomalies.  There are further topics of interest in the 
physics and mathematics of anomalies, that are discussed in detail in the 
references cited for further reading, and elsewhere in the literature.  
We briefly describe a few of them here.
\medskip
\noindent
(1) {\it Anomalies in other spacetime dimensions and in string theory}.
We have focused above on anomalies in four dimensional spacetime, but 
there anomalies of various types both in lower 
dimensional quantum field theories 
(such as theories in 2 and 3 dimensional spacetimes), 
and in quantum field theories in higher dimensional spacetimes 
(such as $N=1$ supergravity in 10 dimensional spacetime). Anomalies also 
play an 
important role in the formulation and consistency of string theory. The 
bosonic string is consistent only in 26 dimensional spacetime, and the 
analogous supersymmetric string is consistent  only in 10 dimensional 
spacetime, because in other dimensions both of these theories 
violate Lorentz invariance after quantization.  
In the Polyakov path integral 
formulation of these string theories, these special dimensions are associated 
with the cancellation of the Weyl anomaly, which is the relevant form 
of the trace anomaly discussed above.  Yang--Mills, gravitational, and mixed 
Yang--Mills gravitational anomalies make an 
appearance both in $N=1$ 10 dimensional supergravity 
and in superstring theory, and 
again special dimensions play a role.  In these theories, only 
when the associated internal symmetry groups are either $SO(32)$ or  
$E_8\times E_8$ is elimination of all anomalies possible, by cancellation   
of hexagon diagram anomalies with anomalous tree diagrams involving 
exchange of a massless antisymmetric two-form field. This mechanism,  
due to  Green and Schwarz, requires the factorization of a sixth 
order trace invariant that 
appears in the hexagon anomaly in terms of lower order invariants, as well 
as two numerical conditions on the adjoint representation generator 
structure, restricting the allowed gauge groups to the two noted above.   
\medskip
\noindent
(2)  {\it Covariant Versus Consistent Anomalies; Descent Equations.}
The non-Abelian anomaly of eqns [4a,b]  is called the ``consistent anomaly'', 
because it obeys the Wess-Zumino consistency conditions of eqn [12c].  
This anomaly, however, is not gauge-covariant, as can be seen from the fact 
that it involves not only the Yang-Mills field strengths $F_{V,A}^{\mu\nu}$, 
but the potentials $V^{\mu},\,A^{\mu}$ as well.  It turns out to be possible, 
by adding appropriate polynomials to the currents, to transform the 
consistent anomaly to a form, called the ``covariant anomaly'', which is 
gauge-covariant under gauge transformations of the potentials 
$V^{\mu},\,A^{\mu}$.  This anomaly, however, does not obey the Wess-Zumino 
consistency conditions, and cannot be obtained from variation of an effective 
action functional. 

The consistent anomalies (but not the covariant anomalies) obey a remarkable 
set of relations, called the Stora-Zumino descent equations, that relate the 
Abelian anomaly in $2n+2$ spacetime dimensions to the non-Abelian anomaly 
in $2n$ spacetime dimensions.  This set of equations has 
been interpreted physically by Callan and Harvey as reflecting 
the fact that the Dirac 
equation has chiral  zero modes in the presence of strings in $2n+2$ 
dimensions and of domain walls in $2n+1$ dimensions.   
\medskip
\noindent
(3)  {\it Anomalies and fermion doubling in lattice gauge theories}.
A longstanding problem in lattice formulations of gauge field theories is 
that when fermions are introduced on the lattice, the process of 
discretization introduces an undesirable doubling of the fermion particle 
modes.  In particular, when one attempts to put chiral gauge theories, such 
as the electroweak theory, on the lattice, one finds that the doublers 
eliminate the chiral anomalies, by cancellation between modes with 
positive and modes with negative axial-vector charge.
  Thus for a long time 
it appeared doubtful whether chiral gauge theories could be simulated on 
the lattice.  However, recent work has led to formulations of 
lattice fermions that use a mathematical analog of a domain wall to 
successfully incorporate chiral fermions, and the chiral anomaly, into 
lattice gauge theory calculations.  
\medskip
\noindent
(4) {\it Relation of anomalies to the Atiyah-Singer index theorem.} 
The singlet ($\lambda_A^a=1$) 
anomaly of eqn [4b] is closely related to the Atiyah-Singer index theorem. 
Specifically, the Euclidean spacetime integral of the singlet anomaly 
constructed from a gauge field 
can be shown to give the {\it index} of the related  Dirac operator for a 
fermion moving in that background gauge field, where the index is 
defined as the  
difference between the numbers of right and left handed zero eigenvalue 
normalizable solutions of the Dirac equation.   Since the index is a 
topological invariant, this again implies that the Euclidean 
spacetime integral of 
the anomaly is a topological invariant, as noted above in our discussion 
of instanton-related applications of anomalies.  
\bigskip
\leftline{\bf Retrospect}
\bigskip
The wide range of implications of anomalies has surprised -- even 
astonished -- the founders of the subject.  New anomaly applications 
have appeared within the last few years, and very likely the future will 
see continued growth of the area of  quantum field theory concerned with the 
physics and mathematics of anomalies.      
\vfill\eject
\leftline{\bf Further Reading}
\bigskip
\noindent
Adler, SL (1969) Axial-Vector Vertex in Spinor Electrodynamics.  
{\it Phys. Rev.} 177: 2426-2438.
\bigskip 
\noindent
Adler, SL (1970) Perturbation Theory Anomalies.  In: Deser, S, M Grisaru, and 
H Pendleton (eds.), Lectures on Elementary Particles and Quantum 
{}Field Theory, vol. 1, pp. 3-164.  M. I. T. Press, Cambridge, MA.\hfill\break 
\bigskip
\noindent
Adler, SL (2004).  Anomalies to all Orders.  ArXiv: hep-th/0405040.  
To appear in:  ' Hooft, G (ed.) Fifty Years of Yang-Mills Theory. 
World Scientific, Singapore.\hfill\break
\bigskip
\noindent
Adler, SL and WA Bardeen (1969)  Absence of Higher Order Corrections in the 
Anomalous Axial-Vector Divergence Equation. {\it Phys. Rev.} 182: 1517-1536.
\hfill\break
\bigskip
\noindent
Bardeen, W (1969)  Anomalous Ward Identities in Spinor Field Theories. 
{\it Phys. Rev.} 184: 1848-1859.\hfill\break
\bigskip
\noindent
Bell, JS and R Jackiw (1969)  A PCAC Puzzle:  $\pi^0 \to \gamma \gamma$ in 
the $\sigma$-Model.  {\it Nuovo Cimento} A 60: 47-61.\hfill\break
\bigskip
\noindent
Bertlmann, RA (1996)  Anomalies in Quantum Field Theory. Clarendon Press, 
Oxford.\hfill\break
\bigskip
\noindent
{}Fujikawa, K and H Suzuki (2004)  Path Integrals and Quantum Anomalies.  
Oxford University Press, Oxford.\hfill\break
\bigskip
\noindent
Green, MB, JH Schwarz, and E Witten (1987).  Superstring Theory, vol. 2, 
secs. 13.3-13.5. Cambridge University Press, Cambridge.\hfill\break   
\bigskip
\noindent 
Hasenfratz, P (2004)  Chiral symmetry on the lattice.  
ArXiv: hep-lat/0406033. 
To appear in:  ' Hooft, G (ed.) Fifty Years of Yang-Mills Theory. 
World Scientific, Singapore.  See also   
M Golterman (2001) Lattice Chiral Gauge Theories, ArXiv: hep-lat/0011027,  
{\it Nucl. Phys. Proc. Suppl.} 94: 189-203; H Neuberger (2000) Chiral 
Fermions on the Lattice,  ArXiv: hep-lat/9909042, {\it Nucl. Phys. 
Proc. Suppl.} 83: 67-76.\hfill\break
\bigskip
\noindent
Jackiw, R (1985)  Field Theoretic Investigations in Current Algebra and   
Topological Investigations in Quantum Gauge Theories. 
In:  Treiman, S, R Jackiw, B Zumino and E Witten (eds.)  Current Algebra 
and Anomalies.    Princeton University Press, Princeton and 
World Scientific, Singapore\hfill\break 
\bigskip
\noindent
Jackiw, R. (2004) Fifty Years of Yang-Mills Theory and Our Moments of 
Triumph. ArXiv:  physics/0403109 (same text, slightly different title).     
To appear in:  ' Hooft, G (ed.) Fifty Years of Yang-Mills Theory. 
World Scientific, Singapore.\hfill\break
\bigskip
\noindent
Makeenko, Y (2002)  Methods of Contemporary Gauge Theory, chapt. 3.  
Cambridge University Press, Cambridge.\hfill\break
\bigskip
\noindent
Polchinski, J (1999) String Theory, vol.1, sec. 3.4 and vol. 2, sec. 12.2. 
Cambridge University Press,  Cambridge.  
\bigskip
\noindent
Shifman, M (1997)  Non-Perturbative Dynamics in Supersymmetric Gauge 
Theories.  ArXiv: hep-th/9704114; 
{\it Prog. Part. Nucl. Phys.} 39:1-116. \hfill\break  
\bigskip
\noindent
van Nieuwenhuizen, P (1988)  Anomalies in Quantum Field Theory: Cancellation  
of Anomalies in d=10 Supergravity.  Leuven University Press, Leuven.  
\hfill\break
\bigskip
\noindent 
Volovik, GE (2003) The Universe in a Helium Droplet, chapt. 18. 
Clarendon Press, Oxford.\hfill\break
\bigskip
\noindent
Weinberg, S (1996)  The Quantum Theory of Fields, vol. II Modern 
Applications, chapt. 22. 
Cambridge University Press, Cambridge. \hfill\break
\bigskip
\noindent 
Zee, A (2003)  Quantum Field Theory in a Nutshell, sec. IV.7.  
Princeton University Press, Princeton and Oxford.\hfill\break
\bigskip
\noindent
\bigskip
\noindent
\bigskip
\noindent
\bigskip
\noindent
\bigskip
\noindent
\bigskip
\noindent
\bigskip
\noindent
\bigskip
\noindent
\vfill
\eject
\leftline{\bf Figure Captions}
\noindent
{\bf Figure 1}   The $AVV$ triangle diagram responsible for  
the Abelian chiral anomaly.  
\bigskip
\noindent
\vfill\eject
\leftline{\bf Keywords}
\bigskip

Anomalies

Axial-vector current  

Chiral anomaly     

Descent equations    

Electroweak theory   

Instantons           

Lattice gauge theory      

Non-Abelian gauge theory

Quantum Electrodynamics

Quantum Chromodynamics

String theory          

Theta vacuum

Trace anomaly           
\bigskip
\bigskip
\leftline{\bf Nomenclature}
\bigskip
MeV   ~~~~~~million electron volts 

\bigskip
\bye